\documentclass[fleqn,12pt]{article}
\usepackage{amsmath}
\usepackage{graphicx}
\usepackage{amsfonts}
\usepackage{amssymb}
\begin{document}
\noindent {\bf Z(3) Symmetric Dimensional Reduction of (2+1)D QCD}

\vspace{0.5cm}  
\noindent P. Bialas$^{1}$, A. Morel$^{2}$, B.Petersson$^{3}$ and K. Petrov$^{3}$\\

\noindent$^{1}$Inst. of Comp. Science, Jagellonian University\\ 
\noindent33-072 Krakow, Poland\\

\noindent $^{2}$Service de Physique Th\'eorique de Saclay, CE-Saclay,\\ 
\noindent F-91191 Gif-sur-Yvette Cedex, France \\

\noindent $^{3}$Fakult\"at f\"ur Physik, Universit\"at Bielefeld\\ 
\noindent P.O.Box 100131, D-33501 Bielefeld, Germany\\

\vspace{0.5cm}

Dimensional reduction is  a  powerful technique 
to study the high-temperature
behaviour of gauge theories. The efficiency of this method was tested in
a detailed investigation of the reduction to 2D of (2+1)D QCD 
\cite{paper1},\cite{paper2}, showing that it is valid down
to temperatures $T$ as low as 1.5 times $T_c$, the critical temperature for
deconfinement in (2+1)D. As it is standard, the reduced model was
derived from a perturbative expansion in powers of the (static) time
component $A_0$ of the gauge field, the non-static components 
being integrated out at one loop order. Such an expansion breaks the $Z_3$
symmetry which the effective model for the Polyakov loops would have.
Hence the reduced model, shown to work in the high
$T$ phase of the full theory, cannot predict confinement as $T$ is lowered.
Here we present
a candidate for a $Z_3$-symmetric reduced action. A similar approach for (3+1)D and possible consequences for QCD were
discussed by Pisarski
\cite{Pisarski:2000eq}.

In the standard reduction, the 2D lattice action has the form \cite{paper1}%
\begin{equation}
\label{old}
S_{eff}^{2}\left(  U,A_{0}\right)  =S_{W}^{2}\left(  U\right)  +S_{int}\left(
U,A_{0}\right)  +S_{pot}\left(  A_{0}\right),
\end{equation}
where $S_{W}^{2}\left(  U\right)  $ is the 2D Wilson action and $S_{int}\left(
U,A_{0}\right)$ the gauge invariant kinetic term for the adjoint
``Higgs'' $A_0$
\begin{equation}
S_{int}\left(  U,A_{0}\right)  =\frac{L_{0}\beta_{3}}{6}\sum_{x}\sum
_{i}Tr\left[  \left(  D_{i}A_{0}\right)  ^{2}\right]
\end{equation}%
\begin{equation}
D_{i}A_{0}=U\left(  x;i\right)  A\left(  x+ai\right)  U\left(  x;i\right)
^{-1}-A\left(  x\right)
\end{equation}

 The local potential is%
\begin{equation}
\label{pot}
S_{pot}\left(  A_{0}\right)  =\sum_{x}h_{2}TrA_{0}^{2}\left(  x\right)
+h_{4}\left(  TrA_{0}^{2}\left(  x\right)  \right)  ^{2}.%
\end{equation}
The couplings are functions of the 3D gauge coupling $g_{3}$, which sets the scale, and temperature $T$:
\begin{equation}
\label{coeff}
h_{2}=-\frac{9\mu}{\pi L_{0}\beta_{3}};\,\,\,h_{4}=\frac{9}{16\pi\beta_{3}^2};\,\,\,%
\beta_{3}=\frac{6}{ag_{3}^{2}};\,\,\,L_{0}=\frac{1}{aT};\,\,\,\mu=  \ln L_{0}+\frac{5}{2}\ln2-1
\end{equation}
Instead of $A_0$, we start with the Polyakov loop $V(x)\equiv
\exp(i\,L_0\,A_0(x))$, whose gauge and $Z_3$-invariant kinetic term
is
\begin{equation}
S_{U,V} \; = \; \frac{\beta_3}{L_0} \;
\sum_{x} \sum_{\mu =1,2}
\biggl( 1 - \frac{1}{3} \Re\, Tr U(x;\mu)
V(x+a\widehat \mu) U(x;\mu)^{-1} V^{-1}(x)
\biggr). 
\end{equation}
We then construct a local potential $S_V$
in $V$, again gauge and $Z_3$-symmetric.  
Since $V$ is an $SU(3)$ matrix, gauge invariance implies that $S_V$ 
depends on two independent variables only at each lattice point, 
$v(x)\equiv 1/3\,Tr\,V(x)$ and $v^{\star}(x)$ for example, whose
two independent and
$Z_3$-invariant combinations are $\vert v(x)\vert ^2$ and $\Re (v^3(x))$.

Since $\langle {v(x)}\rangle$ is an order parameter for $Z_3$, it vanishes in the
symmetric phase, and in this phase we expect that the
effective action may be truncated to the two above lowest order terms in
$v(x)$. We write for it 
\begin{equation} \label{newaction}
S_{Z_3}=S_{W}^{2}\left(  U\right)+S_{U,V}+
\sum_{x}\left(  \lambda_{2}\,|v(x)|^{2}\;+\;\lambda
_{3}\,(v^{3}(x)+v^{\ast3}(x))\right)  .
\end{equation}

Suppose now we move in the {$\lambda_2,\lambda_3$} plane from a point
in the symmetric phase and cross a $Z_3$ breaking transition
line. In the broken phase, we may expand the action in
$A_0$ around zero mod($2\pi/3$) and determine $\lambda_2$ and 
$\lambda_3$ in such a way
the terms of order 2 and 4 match  the action (\ref{old}). The
result is
\begin{equation} 
\label{newcoeff}
\lambda _{2}=\frac{9}{16} \frac{20\mu -1}{\pi L_{0}^{2}};%
\qquad \lambda _{3}= -\frac{3}{8}  \frac{12\mu -1}{\pi L_{0}^{2}}%
\end{equation}

Preliminary results of a numerical investigation with the action
(\ref{newaction},\ref{newcoeff}) indicate that a second order
phase transition does exist. A study of properties of the  
$Z_3$-broken phase is under way, with the aim to compare the
Polyakov loops correlations in both the former and the new model
at high temperature.

\end{document}